# Nematic Ordering Problem As Polymer Excluded Volume One

A.N.Yakunin

Karpov Institute of Physical Chemistry, ul.Vorontsovo Pole 10, Moscow 103064, Russia

Tel:7 (095) 917-4330, FAX:7 (095) 975-2450

E-mail: yakunin@cc.nifhi.ac.ru

**Abstract:** The new method of the mean-field approximation is extended. An approach which enables to estimate some parameters of the transition from the isotropic state of hard sticks to the nematic ordering phase is suggested. An technique of the investigation of the model jointed to the so-called polymer excluded volume problem is proposed. The critical exponents are estimated. The transition of a swelling polymer coil to ideal is revealed as the polymerization degree of the chain increases. The entanglement concentration obtained is in accordance with experimental data for polymers with flexible chains. The number of monomers between neighbor entanglements assumes to be the ratio of the stick length to its diameter. The comparison of the theory with other ones and recent experimental data is carried out.

# 1. INTRODUCTION

In 1966 M. Fisher obtained the expression for the probability density of self-avoiding walks or for the distribution function of polymer chain ends[1]. He assumed that the generating function, $P(\mu, \mathbf{r})$, of the polymer statistics, called also the excluded volume problem, was roughly analogous to the pair correlation function, $G(\mathbf{r}, 1/T)$, of the two-dimensional Ising model in zero field as it followed from Onsager exact solution. Here $\mu$ was the ratio of the chemical potential of a chain monomer to temperature, $T$, which was expressed in energy units; $\mathbf{r}$ was the distance between correlated points.

The subsequent extending of the theory was connected with P.-G. de Gennes[2] who using the Wilson renormalization method[3] showed that the polymer excluded volume problem could be decided in frameworks of the fluctuation theory of phase transitions and critical phenomena. The Hamiltonian of the task was chosen in the form suggested by Ginzburg and Landau for the phenomenological theory of superconductivity.

Our main assumption we will use is to regard the Fisher distribution only as one factor in the expression of a full probability for polymer statistics. The second factor can be taken into account by the way we will demonstrate below. As a result we are going to estimate some parameters of the transition from the isotropic state of hard sticks to the nematic ordering phase such as i) the product of the volume fraction of the sticks and the ratio of the stick length to the diameter, ii) the maximum value of the parameter of the flexibility above that one can not regard the stick as hard. The total method will lead to better understanding both polymer and liquid crystal (LC) problems.

# 2. RESULTS AND DISCUSSION

**A. Pair correlation function of the space density correlation.** We are going to consider the pair correlation function of the space density correlation for a polymer chain with the excluded volume. Our assumption we will use here and below is to regard all of the probability (density) distribution functions as those having a very sharp maximum. In that case we can replace variables by their mean values.

Define the pair correlation function $g(R) = <n(\mathbf{0})n(\mathbf{R}) - <n>^2>/<n>$ for short wave lengths

$$kR > 1 \tag{1}$$

as follows[4]:

$$g(R) = <n>^{-1} | \int (2\pi)^{-d} \exp(i\mathbf{kR}) <n_\mathbf{k}> d^d k |^2. \tag{2}$$

Here and below we suppose that $\mathbf{R}$ is the radius-vector between two monomers located far from each other along the chain; $|\mathbf{R}| = aN^\nu$, that is, a correlation radius $\xi \sim R$ is the single scale distance for the system involved; $\nu$ is the critical exponent[1-3]; a is the monomer diameter (the lattice constant); $\mathbf{k}$ is the wave vector;

$$<n> = N/R^d \tag{3}$$

is the mean concentration of the chain monomers inside the polymer coil; d is the space dimension.

In order to find g(R) it is necessary to define $<n_\mathbf{k}>$. We choose the Debye structural factor, $4d/(ka)^2$, for wave vectors

$$N(ka)^2 >> 1 \tag{4}$$

as a probe function for $<n_\mathbf{k}>$. In order to estimate the integral (2) by a simple way we write the following equation which is true if

$$ka << 1: \tag{5}$$

$$<n_\mathbf{k}> \sim 1/(\exp(\theta)-1) = \sum_{N=1}^{\infty} \exp(-\theta N) \tag{6}$$

where

$$\theta = (ka)^2/(2d). \tag{7}$$

In that case we can obtain the following equation for g(R) by integrating (2) over angles:

$$g(R) \sim a^{-4} R^{-2} \langle n \rangle^{-1} \left| \int_0^\infty \sum_{N=1}^\infty \sin(kR) \, k^{d-2} \exp(-N(ka)^2/(2d)) \, dk \right|^2. \tag{8}$$

We do not write a constant in (8) since the estimation (6) of $\langle n_k \rangle$ we use is not correct when k→0 (see (4) and also (1)). We are going to calculate the constant below.

In (2) we change the order of the integration and the summation, integrate over dk, then replace the summation over N by the integration over dN between the limits from 0 to ∞ since the integrant converges. As a result we find the well-known (if d=3, ν=0.5) expression for the Debye correlation function in the co-ordinate space:

$$g(R) \sim a^{-4} R^{-2} \langle n \rangle^{-1} = a^{-4} R^{d-2}/N = 1/a^2 r \tag{9}$$

where $r \sim aN^{1-\nu}$ and we used (3). However, if ν > 0.5 then r << R since we always regard that N >> 1. It means as we think that there exists an fluctuation attraction between the chain ends owing to screening the volume interactions since the end monomers seem to repulse approximately two times weaker than other monomers. The mechanism of the interaction in polymer solutions was described elsewhere[5]. The attraction compresses the coil in one direction. Consequently, we may assume that its shape looks like an oblate spherical ellipsoid. We can estimate an maximum gyration radius, $r_g$, from the third invariant of the deformation tensor as $R^3 \sim r \, r_g^2$ and find that

$$r_g \sim aN^{2\nu-0.5}. \tag{10}$$

It should be underlined that real polymer chains can be broken by thermal fluctuations if (10) is not satisfied. Speaking otherwise the formula (10) is a condition for the invariant theory.

Now we are going to continue to study properties of the probability density[1]:

$$p_N(r) \cong B/R^d \, (r/R)^\varphi \exp(-(r/R)^{1/(1-\nu)}) \tag{11}$$

where $B^{-1} = (2\pi\nu)^{1/2} (1-\nu)^{1/\nu} \Gamma(1/\nu) S_d$, $S_d = 2\pi^{d/2}/\Gamma(d/2)$ is the surface area of the d-dimensional sphere of unit radius;

$$\varphi = 0.5/(1-\nu) + 1/\nu - d. \qquad (12)$$

We will use the "symmetrical" function (11) although we have found that the coil shape is asymmetric. The fact is caused by the following reason. A priori the function P(0, **r**) is not known. One may suppose that the function is "symmetrical" since the space is isotropic at large scales. If we introduce the two mean scale lengths r and $r_g$ (respectively, parallel and perpendicular to the field compressing the coil) then one can obtain probability density distribution functions as well as the function (11) was received[1]. However, in that case r becomes a function of R, and the exponent $\varphi$ depends on details of defining r and $r_g$. We will use the simplest function (11). We hope to find the rest parameters of the model from the invariants of the deformation tensor. If $\lambda = r_g/R \gg 1$ then the first invariant, $I_1$, is proportional to $2\lambda$ and the second invariant, $I_2$, ~ $\lambda^2 = R/r$ (see (10)).

In (11) we substitute $r_g$ for r (r is called the end-to-end distance by Fisher) because this formula is asymptotically true by definition[1] if the ratio of the radii

$$r/R > 1 \qquad (13)$$

but only $r_g$ can satisfy the condition; the rest parameters such as r and R do not satisfy the condition. Thus we make draw an important conclusion: the Fisher distribution does not correspond to the probability density of the end-to-end distance. The probability density of the end-to-end distance is the function which is proportional to g(R) from (9).

**B. Calculation of the critical exponent of the correlation radius; an thermodynamic consideration.** The logarithm of the function (11) is proportional to $\mu N$ in accordance with the way of the derivation of the function[1]. However, in practice we study the system with the constant concentration of monomers in the coil (N = const) but not with the constant chemical potential of a monomer ($\mu$ = const). This thermodynamic condition results in the "renormalization" of the chemical potential (see, for example[6])

$$\mu \sim N^{\alpha-1} \tag{14}$$

where the critical exponent, $\alpha$, may be expressed as[3]

$$\alpha = 2-\nu d. \tag{15}$$

Substituting (10) for r in (11) and using (14) we find that

$$\alpha = (\nu - 1/2)/(1 - \nu). \tag{16}$$

We determine $\nu$ from (15) and (16). If d = 3 it is equal to $1-6^{-1/2} \approx 0.5918$ (the value of $\nu<1$ should be chosen since the polymer chain can not be longer than its length $\sim$ aN) and is in a good agreement with the value of $\nu \approx 0.5918$ obtained by the second order $\varepsilon$ - expansion[3] if the number of components of an ordering field n = 0. The estimation was also obtained by the author[7] using other considerations. It should be noted that the modern methods[8,9,10] result in different not much values of the critical exponents. However, the discussion of the special question is out of the paper.

**C. Estimation of the critical exponent evaluating the total number of self-avoiding walks.** If we define the square of the ordering parameter about the critical point ($\mu_{cr}$=0 where we suppose one may take R $\sim \xi \rightarrow \infty$) as

$$W^2 = \int p_N(r) \, \delta(r - r_g) \, dr \, / \, (<n> g(R) \, a^d) \sim N^{-\nu+\varphi(\nu-0.5)} \sim N^{-2\beta} \tag{17}$$

then we can obtain the critical exponent, $\beta$:

$$2\beta = \nu - \varphi (\nu - 0.5). \tag{18}$$

Comparing the definition with the conventional expression[3]

$$2\beta = \nu (d - 2 + \eta)$$

we see that

$$\eta = -\varphi (\nu - 0.5) / \nu. \qquad (19)$$

Finally, we can find the exponent, $\gamma$, since[3]

$$\gamma = \nu (2 - \eta). \qquad (20)$$

Substituting (19) for $\eta$ in (20) we obtain

$$\gamma - 1 = 2(\nu - 0.5) - \nu\eta = (\nu - 0.5)(2 + \varphi). \qquad (21)$$

Using (12) and the estimation of $\nu = 1-6^{-1/2}$ we receive that $\gamma \approx 1.1757$ and is in accordance with the value of 1.1758 obtained by the second order $\varepsilon$ - expansion[3] if n = 0.

It should be noted that we define the ordering parameter (17) as the ratio of the two probabilities, $p_N(r_g) R^d / N$ and $g(R) a^d$, that is, $W^2$ is the probability to find the gyration radius, $r_g$, for the swelling coil if the end-to-end distance is r.

We have just modified the Fisher method. We ought only to underline that g(R), by its definition[4], satisfies the condition not to locate the different monomers in one site of the lattice and to write the meaning of $\eta \approx 0.0132$ using (19). The $\varepsilon$-expansion results in $\eta \approx 1/64 \approx 0.0156$[2].

**D. Another definition of g(R).** We may substitute (7) for $\theta$ in (6) and, using $\langle n_k \rangle \sim 1/\theta$ when $\theta \to 0$, obtain $g_N(R) = g(R)/N$ for $2 < d < 4$ by fulfilling the procedures we have used above:

$$g_N(R) = (2C') a^{-4} R^{-2} \langle n \rangle^{-1} \left| \int_q^\infty dk \, \sin(kR) \, k^{d-4} \right|^2$$

where qR > 1. Integrating by parts and neglecting the term with the greatest degree of $R^{-1}$ we find

$$g_N(R) = C' a^{-4} R^{-4} \langle n \rangle^{-1} q^{2(d-4)}. \qquad (22)$$

where

$$C' = S_d^2 (2d)^2 / 2(2\pi)^{2d}. \tag{23}$$

If the maximum of $r_g$ is expressed as the reciprocal value of the minimum of the wave vector $1/(aq) \sim r_g/R$ then $aq \sim N^{0.5-v}$ and $g_N(R)$ in (22) is proportional to $g(R)/N$. The value of $q$ satisfies the conditions (1), (4) and (5). Consequently, only now we can try to estimate the constant in (8).

It should be noted that the factor $(2\pi)^{-2d}$ in (23) origins from the density of states in the momentum space. This is in a contradiction with the definition $W^2$ as a condition probability. In that case we must go to the probability from the probability density. Therefore, we sum over all states of a wave vector. Then the following equation for the adjusting (see below) constant, C, of the functions, $g(R)$ and $g_N(R)$, can be written:

$$C = S_d^2 (2d)^2 / 2. \tag{24}$$

## 3. CONCLUSIONS: ESTIMATIONS OF THE ENTANGLEMENT CONCENTRATION AND PARAMETERS OF TRANSITION TO LC STATE

Since we defined above $g(R) a^3 = g_N(R) N a^3$ as a probability it can not be > 1. Let us regard $g(R) a^3 / N$ as a probability of the binary collision of the chain monomers. We may determine $N_e$ from the expression by help of (9), (22), (24) and obtain

$$N_e = C^{1/(2-v)} \approx 283. \tag{25}$$

It means that in generally for low enough values of polymerization degree the probability to interact two different end monomers of the chain is equal to 1 if two-particle collision takes place.

We regard $N_e$ as a value which can correspond to the number of the monomer between of (self-) entanglements of the chain. The reason for this is following. We know that the parameter, $N_e$, is used in studying both the high frequency elastic responses of polymer melt and the low frequency viscous properties. The behavior is possible if $N_e$ is an thermodynamic parameter such as temperature or pressure. Therefore, it must be certainly defined in our theory.

Experimental data show[6,11] that $N_e$ = 50-300. The fact is the first argument for the use of the constant (24).

The second argument can be obtained from following considerations. For high enough N we suppose that the system shall become unstable. Then the transition to an ideal coil will occur and $W^2$ will not depend on N. If $W_{cr}^2 = (B/C)^2$ then $W=W_{cr}$ at $N_{cr} = (C/B)^{1/(2\beta)} \approx 8\times10^6$. On the other hand, g (R) $a^3$ can not be > 1. Consequently, the formula $\ln C = (1 - \nu) \ln N$ gives the value of $N_{cr}$ as for an ideal coil if $\nu = 0.5$: $N_{cr} = C^2 \approx 8\times10^6$. Thus the chain screens its volume interactions and transforms from the swelling state to an ideal coil at $N_{cr}$.

At last, we may obtain the parameters of transition to LC state. The value $N_e$ is received from the determination of the probability of the binary collision of the chain monomers if the two monomers do not occupy the same site of the lattice. At $N < N_e$ the probability is one by its definition. We suppose that sticks with the length-diameter ratio, L/D, proportional to $N_e$ can form the nematic ordering phase since the viscosity of polymer melt is proportional to N in this case. The volume fraction of the sticks may find from (3) as $\Phi_{nematic} = <n>a^3 = N_e^{1-\nu d}$. The product of the value and $N_e$ is

$$\Phi = N_e^{\alpha} = 3.56$$

and must be compared with the value $\Phi_{nematic}$ L/D = 4.5 from Onsager theory[12] or with other models[5].

According to our consideration $N_e$ is an parameter of the flexibility and the value L/D = $N_e$ = 283 is the maximum above that one can not regard the stick as hard.

Thus, we can see that the estimations for the LC tasks may be obtained from the other field of statistical physics. This is the important result of the present work since the suggested method enables us to solve problems in various fields of statistical physics.

The recent SANS experimental data[13] confirm the mentioned above. At low temperatures the exponents of the scaling laws of the dependence of gyration radii on polymer molecular weight in the directions parallel and perpendicular to the director of studied nematic phase are measured: $\nu_\parallel \approx 0.66 \pm 0.02$ and $\nu_\perp \approx 0.46 \pm 0.02$. These values are close to polymeric exponents: $2\nu-0.5$ and $1-\nu$. As the temperature increases $\nu_\parallel$ and $\nu_\perp$ tend to 0.5 when the nematic – isotropic phase transition occurs. The similar effect has been described above for the polymer system: the polymerization degree increase is the prime cause for the transition to the isotropic state.

**ACKNOWLEDGEMENTS**

The author is thankful to Russian Foundation for Basic Research (Grant No 01-03-32225) for financial support.